\documentclass[
aps
,prb,onecolumn
,superscriptaddress
,showkeys
,longbibliography
]{revtex4-2}
\usepackage{graphicx}
\usepackage{dcolumn}
\usepackage{bm}
\usepackage{latexsym}
\usepackage{booktabs,threeparttable}
\usepackage{makeidx}
\usepackage{amsmath,amsfonts,amsthm,amssymb}
\usepackage{nccbbb}%
\usepackage{mathrsfs}
\providecommand{\U}[1]{\protect\rule{.1in}{.1in}}

\usepackage{xcolor}

\usepackage{cleveref}
\crefname{equation}{Eq.}{Eqs.}
\usepackage{natbib}
\usepackage{subfigure}
\usepackage{graphicx}

\usepackage{braket}

\begin{document}

\title{Giant Conductance Enhancement in Graphene Nanoscrolls via Interlayer-Coupled States}

\author{Jia-Cheng Li}
\affiliation{Center for Quantum Frontiers of Research and Technology (QFort), National Cheng Kung University, Tainan 70101, Taiwan}
\affiliation{Program on Key Materials, Academy of Innovative Semiconductor and Sustainable Manufacturing, National Cheng Kung University, Tainan 70101, Taiwan}

\author{Ying-Je Lee}
\affiliation{Center for Quantum Frontiers of Research and Technology (QFort), National Cheng Kung University, Tainan 70101, Taiwan}
\affiliation{Department of Physics, National Cheng Kung University, Tainan 70101, Taiwan}

\author{Yu An Cheng}
\affiliation{Center for Quantum Frontiers of Research and Technology (QFort), National Cheng Kung University, Tainan 70101, Taiwan}
\affiliation{Department of Physics, National Cheng Kung University, Tainan 70101, Taiwan}

\author{Yu-Hsiang Fu}
\affiliation{Research Center for Applied Sciences, Academia Sinica, Taipei 115201, Taiwan}

\author{Xuan-Fu Huang}
\affiliation{Center for Quantum Frontiers of Research and Technology (QFort), National Cheng Kung University, Tainan 70101, Taiwan}
\affiliation{Department of Physics, National Cheng Kung University, Tainan 70101, Taiwan}

\author{Yu-Jie Zhong}
\affiliation{Center for Quantum Frontiers of Research and Technology (QFort), National Cheng Kung University, Tainan 70101, Taiwan}
\affiliation{Department of Physics, National Cheng Kung University, Tainan 70101, Taiwan}

\author{Ion Cosma Fulga}
\affiliation{Institute for Theoretical Solid State Physics, IFW Dresden, Helmholtzstraße 20, 01069 Dresden, Germany}
\affiliation{W\"{u}rzburg-Dresden Cluster of Excellence ct.qmat, Germany}

\author{Carmine Ortix}
\affiliation{Dipartimento di Fisica ``E. R. Caianiello'', Universit\`a di Salerno I-84084 Fisciano (Salerno), Italy}

\author{Ching-Hao Chang}\email{cutygo@phys.ncku.edu.tw}
\affiliation{Center for Quantum Frontiers of Research and Technology (QFort), National Cheng Kung University, Tainan 70101, Taiwan}
\affiliation{Program on Key Materials, Academy of Innovative Semiconductor and Sustainable Manufacturing, National Cheng Kung University, Tainan 70101, Taiwan}
\affiliation{Department of Physics, National Cheng Kung University, Tainan 70101, Taiwan}

\date{\today} 

\begin{abstract}
Graphene nanoscrolls form a unique radial superlattice whose geometry enables transport behaviors distinct from their flat counterparts. We theoretically demonstrate that rolling a graphene nanoribbon into a few-turn nanoscroll (1.0–1.2 turns) triggers a giant enhancement in longitudinal conductance by an order of magnitude (nearly twentyfold). We show that this effect stems from interlayer hopping between open boundaries, which not only transforms localized edge states into one-dimensional-like conducting channels and also induces a significant Fermi energy shift. Our results reveal an inverse scaling of conductance with turn number, establishing GNSs as a platform for geometry-driven quantum transport and offering a tunable approach to engineering nanoarchitectural electronic states.
\end{abstract}

\maketitle

\section{Introduction}

Efficient external control of device conductance is the key to modern electronic technology. Common paradigms include the electrostatic modulation of drift currents in p-n junctions and the magnetic field-driven reorganization of magnetic textures in magnetoresistive systems. While these strategies effectively manipulate semiclassical charge transport, achieving not only qualitative but also quantitative control over quantum transport will be a cornerstone for next-generation quantum technologies.

Recently, the emergence of \textit{nanoarchitectonics}, where the physical geometry and assembly of a system dictate its electronic properties, has opened new avenues for transport engineering. A prominent example is the discovery of \textit{twistronics}, where the relative rotation between layers in van der Waals heterostructures induces flat bands and correlated phases, such as superconductivity in twisted bilayer graphene~\cite{Cao2018}. Beyond twisting, the rolling of 2D materials into \textit{radial superlattices}—specifically graphene nanoscrolls (GNSs)—offers a unique platform where topology, number of turns, and periodic boundary overlaps modulate the dimensionality and distribution of quantum states~\cite{Xie2009, cui2018, Zhao2021, Zhong2022, Wang2023Aug}. Recent experimental and theoretical work highlights that interlayer coupling dominates the electronic landscape of GNSs \cite{Zhang2025, Zhong2025Mar}. Consequently, the spin-polarized currents and remarkable magnetoconductance can be actively driven by tailoring the inter-turn chirality and the total turn number\cite{Zhang2025, Zhong2025Mar}, respectively.

In this work, we theoretically demonstrate that the longitudinal quantum conductance of a graphene nanoribbon can be directly enhanced by over an order of magnitude (nearly twentyfold) by rolling it into a few-turn nanoscroll, which can be experimentally realized \cite{Schmidt_2018}. This dramatic enhancement occurs specifically in the 1.0–1.2 turn regime. We establish that this conductance jump stems from a topological transition in the electronic manifold: as the ribbon is rolled, interlayer coupling between the overlapping boundaries acts as a non-trivial perturbation that lifts the degeneracy of the localized edge states. This transformation not only provides a high-velocity pathway for charge carriers but also triggers a self-consistent shift in the Fermi level at constant density.  Our findings suggest that the nanoscroll turn number acts as a tuning knob for the nature of quantum states and quantum transport (see Fig.1(c) and Fig.2), offering a route toward geometry-driven nanoelectronics.

\begin{figure}[h!]
\includegraphics[scale=0.55]{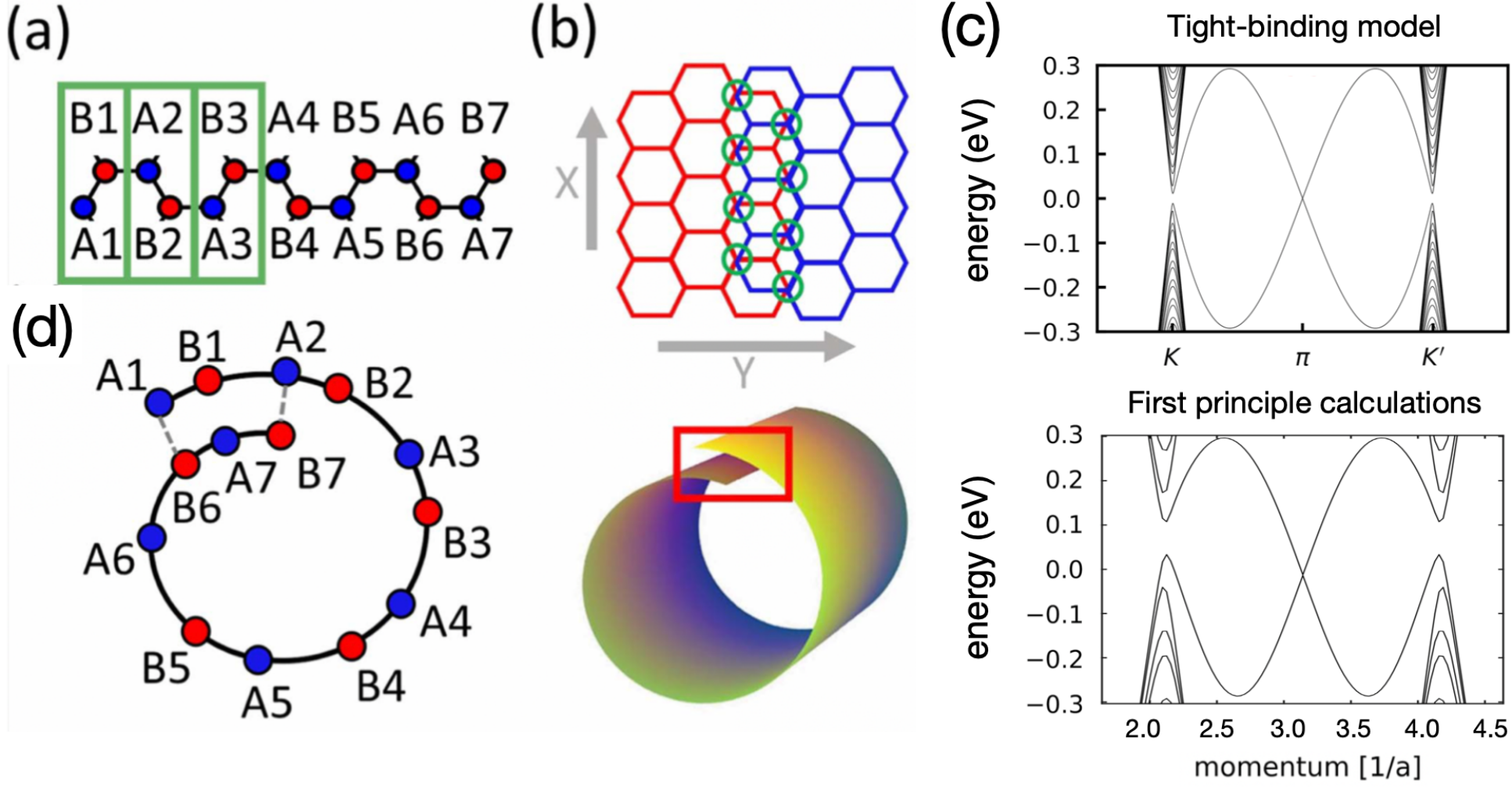}
\caption{Figure panels: (a) Unit atom chain of the system; (b) three-dimensional spiral model with a red frame indicating the boundary green circle highlighting the zigzag interlayer boundary condition; (c) band structures of a 1.0-turn nanoscroll calculated using the tight-binding model with an arc length of 99.5 nm ($m=467$) and the DFT model with an arc length of 21.59 nm; (d) cross-section of the carbon nanoscroll shows the AB stacking between interlayers. The band structure of the carbon nanotube constructed from the same number of atoms is provided in the Supplemental Material.} 
\label{one_turn_nanoscroll_fig1}
\end{figure}

\begin{figure}[h!]
\includegraphics[scale=0.42]{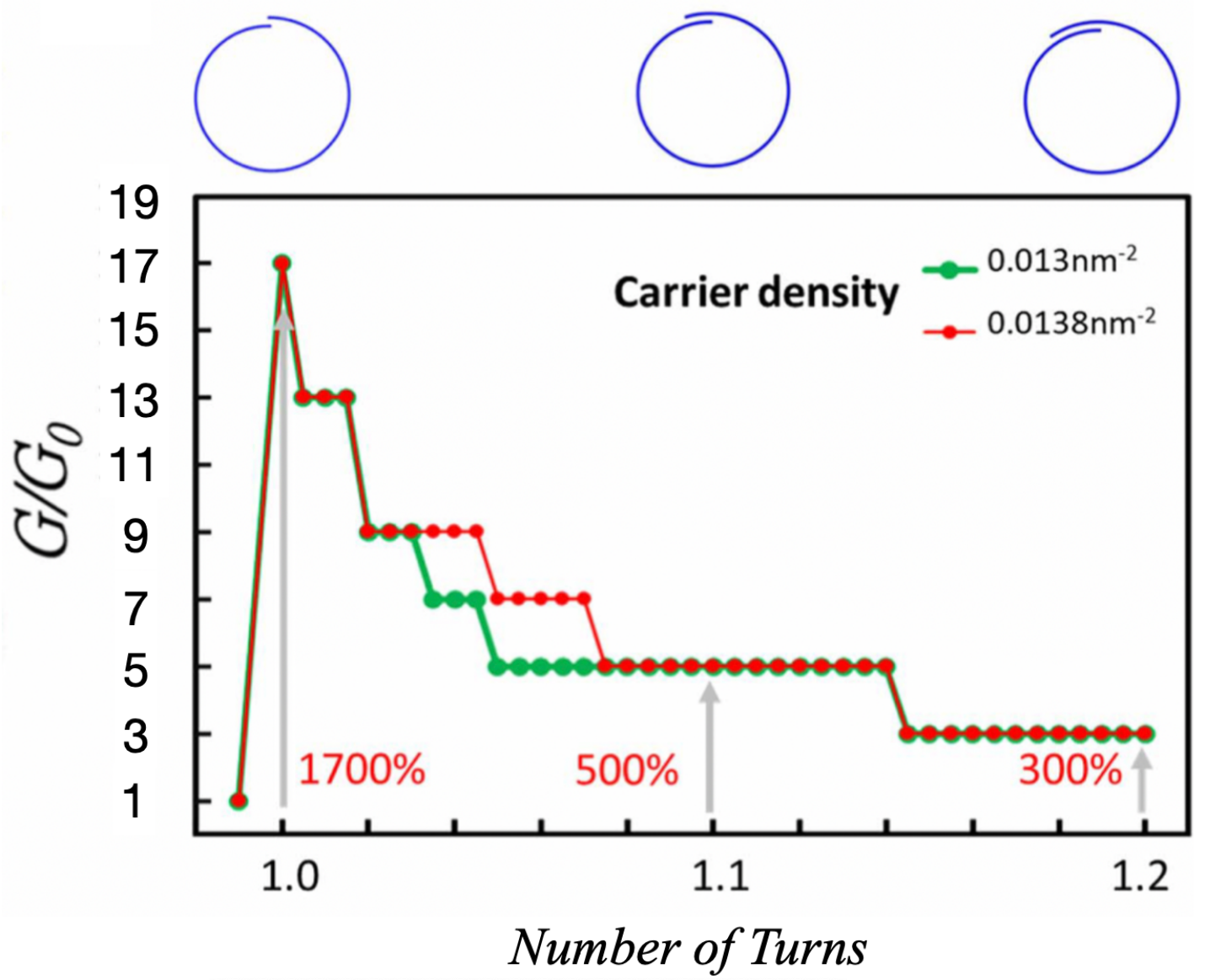}
\caption{ Three panels on the top of figure presents cross-sections of a 99.5-nm-long GNS with 1.0, 1.1, and 1.2 turns. The main figure illustrates the relationship between the number of turns and conductance at carrier densities of 0.0138 nm$^{-2}$ and 0.013 nm$^{-2}$, demonstrating a significant increase in conductance with additional turns.}
\label{one_turn_nanoscroll_fig2}
\end{figure}

\section{Theoretical Method} 

\textit{Model} --
We employ a tight-binding approach to analyze GNS \cite{lee2025,fu2026,hsu2026}. 
The complete Hamiltonian \( H \) is constructed using the basis \(\{A_1, B_1; A_2, B_2; \ldots; A_{2N-1}, B_{2N-1}\}\) to form a unit cell, while also considering the zigzag-edge boundaries to account for interlayer electronic tunneling. Here, \( A \) and \( B \) represent the A- and B-type carbon atoms, and the numerical subscripts indicate the pairs of carbon atoms: \( A \)-\( B \) in graphene, as illustrated in Fig.~\ref{one_turn_nanoscroll_fig1}.
We solve for the quantum states of the nanoscroll with varying numbers of turns and calculate the conductance corresponding to these turns under specific carrier densities. 
To validate the results obtained from the tight-binding model, we compare its calculated band structure with that obtained from DFT calculations.

We begin with the spinless tight-binding model of a zigzag-type graphene nanoribbon with nearest-neighbor hopping
\begin{equation}
H_{\rm rib} = \gamma_0 \sum\limits_{\langle r,r' \rangle} a^{\dagger}_{r} b_{r'} + h.c.,
\end{equation}
where $\gamma_0$ represents the intralayer coupling strength, and the electronic operators $a$ and $b$ denote the annihilation operators for A-site and B-site sublattices, respectively. Here, $r = (i, j)$, with $i$ indicating the $i^{\rm th}$ unit cell along the axial direction, and $j$ representing the $j^{\rm th}$ AB-pair within the ribbon's open boundaries. A nanoribbon with a width of $m$ unit cells is simulated by imposing the condition $j \leq m$. 
%

When the nanoribbon rolls up into an $n$-turn GNS, the full Hamiltonian $H$ can be expressed as $H=H_{rib}+H_{inter}$, where
\begin{equation}
    H_{\rm inter}=\gamma_1 \sum\limits_{j=1}^{j_{\rm max}}a^{\dagger}_{i,j} b_{i,k}+h.c..
\end{equation}
This equation accounts for the interlayer A-B site hopping (AB-stacking) within the nanoscrolls. The interlayer hopping strength is denoted by $ \gamma_1 $.
 The net number of interlayer hopping is given by $j_{\rm max}=m+1-2\lfloor m/{2N_{\rm turn}}\rfloor$ and $k=j+2\lfloor m/{2N_{\rm turn}}\rfloor-1$, where the number of unit cells ($m$) is odd in this case and $ N_{\text{\rm turn}} $ represents the number of turns of GNS.
For example, a 1.0-turn GNS with 467 AB pairs, as shown in Fig.~\ref{one_turn_nanoscroll_fig1}(b), is modeled by setting $N_{\mathrm{turn}}=1$, $m=467$, and $j_{\max}=2$. We note that in a 1.0-turn GNS, the number of sites with interlayer coupling within a unit cell can be reduced from two (see Fig.~\ref{one_turn_nanoscroll_fig1}(b)) to one by changing the total number of AB pairs from an odd $(m=467)$ to an even $(m=466)$ value, resulting in a  different GNS ground state~\cite{lee2025}. 
However, the giant conductance enhancement shown in Fig.2 remains valid for the GNS with even $m$ because the remarkable Fermi-level shift shown in Fig.4 occurs for nanoscrolls with either an odd or even number of AB pairs (see Supplemental Material).

\textit{Conductance}--
We evaluate the conductance by utilizing the Landauer formula~\cite{Landauer1957Jul,Bagwell1989Jul,Chang2014Nov,Zhong2022} for the ballistic scattering process
\begin{equation}
G(E_F, T) = \int_{-\infty}^{\infty} G(E, 0) \left.\frac{\partial f}{\partial E}\right|_{E_F} dE,
\end{equation}
where $f$ represents the Fermi-Dirac distribution function and $E_F$ is the Fermi energy. The zero-temperature conductance of our one-dimensional nanostructure, which is perfectly ballistic, is proportional to the number of modes ($N_s$), given by
\begin{equation}
G(E, 0) = \frac{2 e^2 N_s}{h}. 
\label{G0}
\end{equation}
All numerical results presented in this work are calculated at $T = 0$K.


\textit{Numerical Simulation}--
We simulate the GNS system using a tight-binding model with the following parameters: the intralayer coupling between sites A and B is $\gamma_{0}=3.16$~eV~\cite{Kuzmenko2009Oct, McCann2013, PhysRevB.79.165431}, and the interlayer coupling between site A2 (the A site in the second layer) and site B1 (the B site in the first layer) is $\gamma_{1}=-0.381$~eV. The hamiltonian of GNS is constructed by using the Kwant package~\cite{Groth2014Jun} and calculated by Eq.(\ref{G0}), simulating a system with an arc length of 99.5~nm. The results are found to be both qualitatively and quantitatively consistent for different arc lengths, including 50~nm (see the Supplemental Material). To investigate the dependence of carrier properties on the number of turns in the nanoscroll, we calculate the density of states $\rho(E)$ and
the carrier density is given by $n=\int_{0}^{E_F}\rho(E)\,dE$, where $E$ denotes the eigenvalues of the full Hamiltonian $H$ \cite{zenodo}.

\section{Results and discussions}

We model a zigzag graphene nanoribbon within a tight-binding framework, focusing on an odd-AB-pair configuration. As illustrated in Fig.~\ref{one_turn_nanoscroll_fig1}(a), this toy model consists of seven AB pairs. To form the nanoscroll, we consider this ribbon with edges parallel to the $y$-axis, which is then rolled around the $x$-axis for one complete revolution, resulting in the AB-stacked morphology shown in Fig.~\ref{one_turn_nanoscroll_fig1}(d). The atomic arrangement at the scroll boundary is detailed in Fig.~\ref{one_turn_nanoscroll_fig1}(b); edge atoms participate in two intralayer hoppings and one interlayer hopping (green circles), thereby preserving local AB stacking throughout the overlap region. 

When the graphene nanoribbon rolls up into a 1.0-turn nanoscroll, both the TB and DFT calculations show an evolution of the ground-state band structure from flat bands to a band crossing as shown in Fig.~\ref{one_turn_nanoscroll_fig1}(c). Moreover, both the slope and energy range of the band crossing in the 1.0-turn nanoscroll agree well between two calculations. The only difference is that a mild particle-hole asymmetry is observed in the DFT band structure, because the DFT calculation accounts for additional system details, including higher-order hopping terms, breaking the symmetry. The detailed results of DFT calculation are provided in the Supplemental Material.

\begin{figure}[h!]
\includegraphics[scale=0.5]{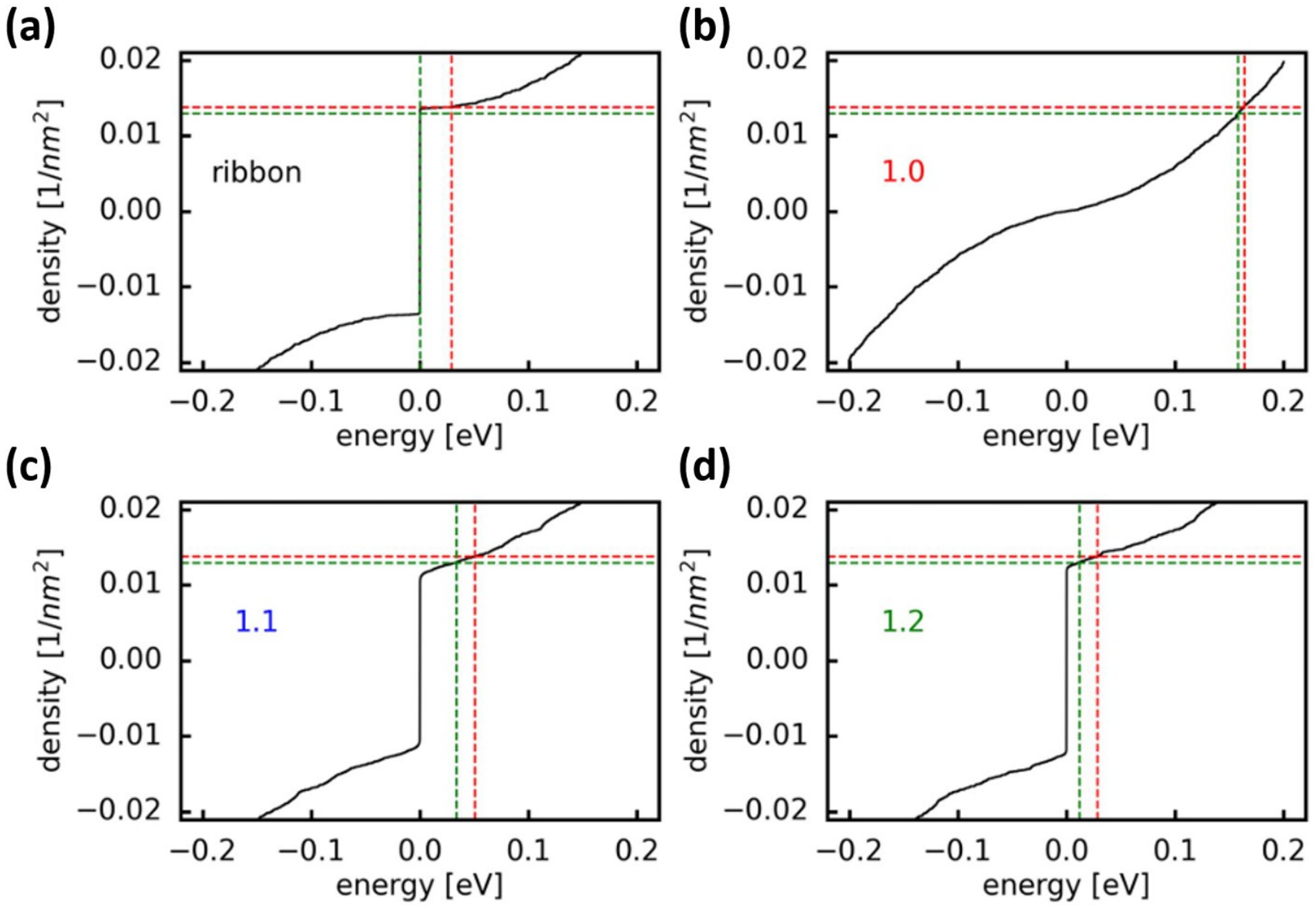}
\caption{The carrier density distribution is presented for (a) ribbon, (b) 1.0-turn, (c) 1.1-turn, and (d) 1.2-turn carbon nanoscrolls. At 1.0 turn, the slope of the carrier density curve decreases significantly, resulting in a smoother variation with energy. As the number of turns increases to 1.1 and 1.2, this trend becomes more pronounced.} 
\label{one_turn_nanoscroll_fig3}
\end{figure}

The main result of this study—the simulated longitudinal quantum conductance of nanoscrolls with an arc length of 99.5 nm—is presented in Fig.~\ref{one_turn_nanoscroll_fig2}, which compares the cross-sectional configurations of nanoscrolls with 1.0, 1.1, and 1.2 turns at fixed arc length. Notably, the one-turn nanoscroll ($N_{\text{turn}} = 1.0$) exhibits a quantum conductance enhancement of up to 1700\% relative to a flat ribbon at carrier densities of 0.0138 nm$^{-2}$ and 0.013 nm$^{-2}$. Although the conductance decreases in discrete steps as the number of turns increases, it remains 300–500\% higher than that of the flat ribbon even at $N_{\text{turn}} = 1.2$.
These specific carrier densities ($n=0.013$ and $0.0138$ nm$^{-2}$) were selected because they correspond to the Fermi levels at the base and the upper edge of the first subband in the flat nanoribbon, respectively (Fig.~\ref{one_turn_nanoscroll_fig3}(a)). These regimes serve as clear reference points for monitoring the evolution of the Fermi level as the ribbon is incrementally rolled into nanoscrolls with varying turn numbers.

The carrier densities as a function of energy for the flat nanoribbon and the 1.0-, 1.1-, and 1.2-turn nanoscrolls are compared in Figs.~\ref{one_turn_nanoscroll_fig3}(a)–(d). In the flat nanoribbon, the carrier density exhibits a nearly vertical increase [Fig.~\ref{one_turn_nanoscroll_fig3}(a)], indicative of a high density of states near the zero energy. However, upon rolling the ribbon into a nanoscroll, this rapid rise is significantly suppressed; the energy dependence near the neutrality point flattens, as seen in Fig.~\ref{one_turn_nanoscroll_fig3}(b). This stark reduction in slope causes a sudden jump in the Fermi level at a fixed carrier density—a shift clearly tracked by the evolution of the green and red reference lines across Figs.~\ref{one_turn_nanoscroll_fig3}(a) and (b), which correspond to carrier densities of $n=0.013$ and $0.0138$ nm$^{-2}$, respectively. Consequently, the geometric transition from a flat ribbon to a 1.0-turn nanoscroll produces a significant reduction in the slope of the carrier density curve alongside an upward shift in the Fermi energy. As the number of turns increases beyond one, the steep slope of the carrier density begins to recover while the Fermi energy decreases, rendering the Fermi-level position highly sensitive to the specific nanoscroll turn number (see Figs.~\ref{one_turn_nanoscroll_fig3}(b), (c) and (d)).


\begin{figure}[h!]
\includegraphics[scale=0.45]{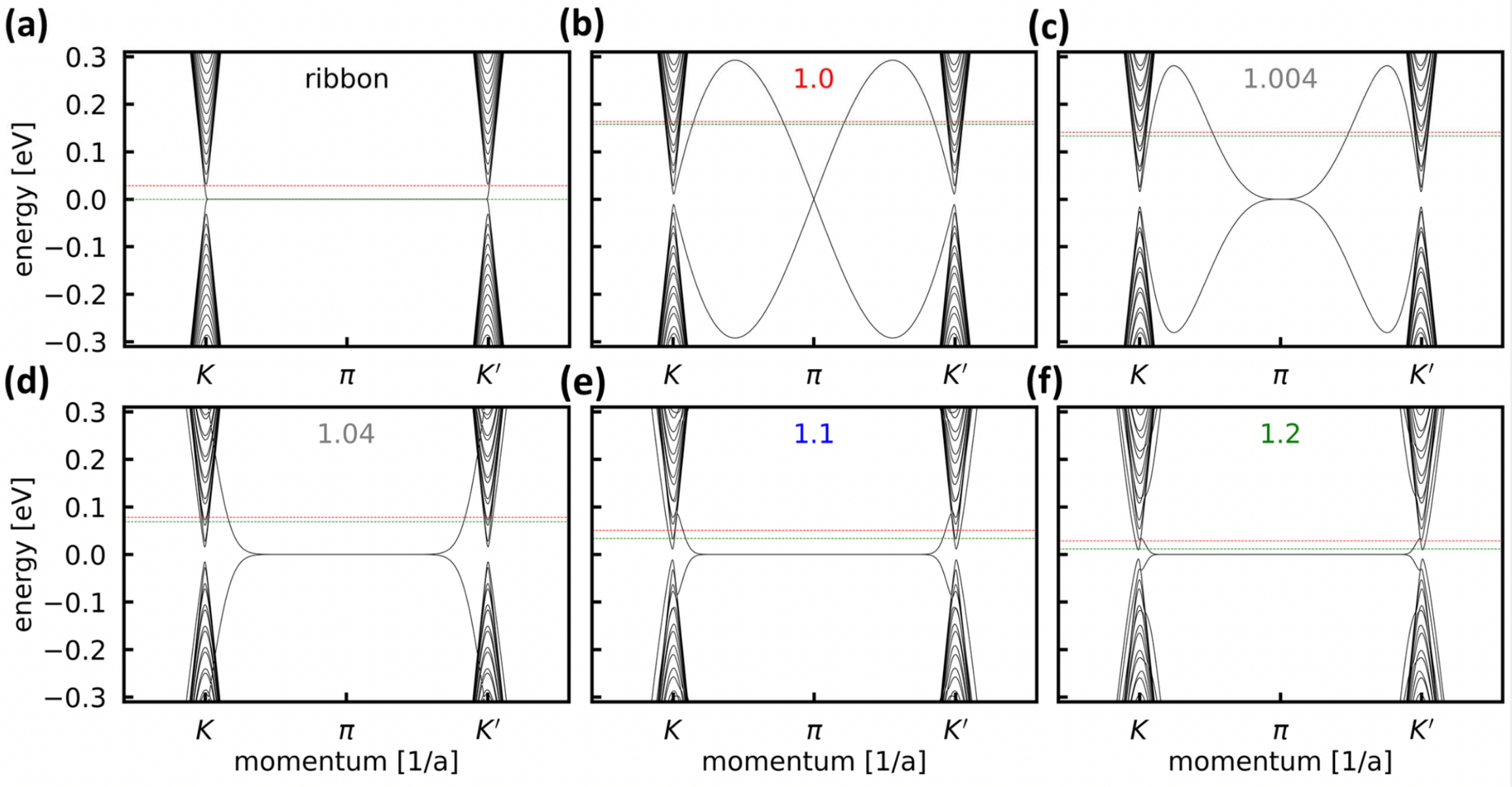}
\caption{The figure illustrates the evolution of the energy band structure from a ribbon to a 1.2-turn nanoscroll. The green and red lines indicate the Fermi energy at carrier densities of 0.013 nm$^{-2}$ and 0.0138 nm$^{-2}$, respectively.}
\label{one_turn_nanoscroll_fig4}
\end{figure}
To understand the origin of the sharp Fermi-level jump at one turn and its subsequent decline at higher turn numbers, Fig.~\ref{one_turn_nanoscroll_fig4} compares the band structures of the flat graphene ribbon against those of nanoscrolls with varying turns. In these plots, the red and green dashed lines indicate the Fermi levels corresponding to the two previously discussed carrier densities. Upon transformation from a ribbon into the 1.0-turn carbon nanoscroll, the ground states at two edges directly hybridize through the interlayer tunneling, leading to the initially flat bands of edge states merge into a band crossing. This transition of band dispersion dramatically reduces the density of states (DOS) at low energies, leading to the sharp jump of Fermi level.
As the turn number of the GNS increases, this hybridization quickly weakens as the overlapping region of the localized edge states decreases, and the Fermi level shifts to lower energy (see Figs.~4(b) to (f)). 

According to the Landauer formula, this elevated Fermi level intersects multiple transport modes, which leads to the pronounced increase in quantum conductance observed Fig.~\ref{one_turn_nanoscroll_fig2}. With an increasing number of turns, the flat band, contributed by the graphene edge states, gradually reappears at low energy, enabling the carrier density to be maintained without a substantial increase in the Fermi level. This process results in fewer conducting modes and a gradual reduction in conductance enhancement.  Although the giant conductance enhancement in a few-turn GNS (Fig. 2) is mainly attributed to the sharp jump in the Fermi level at a fixed carrier density (Fig. 4), a remarkable conductance enhancement of up to 500\% also occurs at a fixed Fermi level of $0.031$ eV, as shown in the Supplemental Material. This enhancement combines the evolution of the ground-state bands from flat bands to a band crossing (adding 4 conduction channels) and the decrease in the energy of the first excited states to below the fixed Fermi level (adding 4 conduction channels), which increases the total number of conduction channels from 2 to 10.

\section{Mechanism Illustration}

To clarify why the ground-state band transitions from a flat to a band-crossing profile upon the formation of a 1.0-turn carbon nanoscroll, we developed a continuum model to describe the interlayer coupling of graphene edge states. The effective Hamiltonian for these edge states is represented as:
\begin{align}
H=
\begin{bmatrix}
   0 & 2\gamma_1 \cos(k_x d)\\
    2\gamma_1 \cos(k_x d) & 0\\
\end{bmatrix},
\end{align}
where the diagonal terms correspond to the zero-energy edge states of the ribbon, and the off-diagonal terms describe the interlayer coupling between edge states in the nanoscroll. Here, $k_x$ is the wave vector along the nanoscroll axis, and $d$ is the interlayer spacing (see Fig.~\ref{one_turn_nanoscroll_fig1}(b)); a detailed derivation is provided in the Supplemental Material). 

The resulting eigenvalues of effective Hamiltonitan, $E = \pm 2\gamma_1 \cos(k_x d)$, qualitatively and quantitatively reproduce the band structure shown in Fig.~\ref{one_turn_nanoscroll_fig4}(b). Our analysis confirms that these band-crossing states originate from interlayer coupling. This coupling drives the giant conductance enhancement exhibited by few-turn scrolls (Fig.~\ref{one_turn_nanoscroll_fig2}) by triggering a sharp shift in the Fermi level and effectively activating multiple conduction channels.


%

\section{Summary}

\textit{In sum}, we combine numerical simulations, first-principles calculations, and analytical calculation to thoroughly investigate the unique band structure and transport characteristics of carbon nanoscrolls. We find that rolling a graphene nanoribbon into a nanoscroll introduces an additional structural degree of freedom that can modify its electronic structure, thereby directly impacting its transport properties.  Our findings reveal that the observed conductance anomaly stems from changes in the DOS, which result in a shift of the Fermi level at a fixed carrier density. Notably, at low carrier concentrations, the transformation from a graphene nanoribbon to a GNS significantly elevates the Fermi level, enabling the system to access three to seventeen times more transport modes compared to a flat configuration. This impressive enhancement is attributed to geometry-induced electronic reconstruction, where the interlayer overlap fundamentally alters the local electronic potential, generating additional conduction channels. Moreover, this conductance enhancement can be observed at temperatures up to $T \lesssim 40$ K, and the observable temperature range can be extended to $T \lesssim 72$ K by reducing the arc length of the GNS from 100 nm to 50 nm \cite{note1}.

Moreover, as the number of turns in the nanoscroll increases, we see a gradual recovery of the DOS near the Fermi level, which subsequently reduces the number of available modes and leads to a stepwise decrease in conductance. This behavior clearly establishes a strong correlation between scroll morphology, energy dispersion, and quantum transport. The agreement between the tight-binding, first-principles, and analytical calculations demonstrates that the surge in conductance directly reflects band evolution driven by the geometry of the scroll, independent of extrinsic effects. Our results unveil a fundamental mechanism through which curvature and interlayer coupling collaborate to precisely tune charge transport in low-dimensional carbon systems, bridging the concepts of quantum interference and geometry-driven electronic design with confidence.

 \section*{Author Contributions} 
J.-C. Li, Y.-J. Lee and Y. A. Cheng contributed equally to this work.

\section*{Acknowledgements}
We acknowledge the financial support by the National Science and Technology Council (Grant numbers 112-2112-M-006-026-, 112-2112-M-004-007 and 112-2112-M-006-015-MY2) and National Center for High-performance Computing for providing computational and storage resources. C.-H. C. thanks A. Huang for helpful discussions and assistance. This work was supported in part by the Higher Education Sprout Project, Ministry of Education to the Headquarters of University Advancement at the National Cheng Kung University (NCKU).


\begin{thebibliography}{21}
\bibitem [{not()}]{note1}%
  \BibitemOpen
  \href@noop {} {}\bibinfo {note} {The relevant temperature scale can be
  estimated by comparing the thermal energy $k_BT$ with the characteristic
  subband spacing. From Fig. 3(a), the first excited-state energy is
  approximately $0.031$ ~eV, corresponding to a characteristic temperature of
  $\frac{\Delta E}{k_B} \approx 360K$. Thus, as long as $k_B T\ll\Delta E$,
  thermal broadening should not significantly smear the subband structure
  responsible for the predicted transport behavior. Using $k_B
  T\lesssim0.1\Delta E$ as a conservative criterion, we estimate that the
  conductance enhancement proposed in our work can be measured at temperatures
  of $T \lesssim 40 K$. Furthermore, when the arc length is reduced from 100 nm
  to 50 ~nm, the first excited-state energy increases to approximately 0.062
  ~eV, corresponding to $\frac{\Delta E}{k_B} \approx 720 K$, and the estimated
  observable temperature range increases to $T \lesssim 72 K$.}\BibitemShut
  {Stop}%
\end{thebibliography}
%
\end{document}